# Dirac Line-nodes and Effect of Spin-orbit Coupling in Non-symmorphic Critical Semimetal *M*SiS (*M*=Hf, Zr)

C. Chen[1]*, X. Xu[2]*, J. Jiang[3,4,5], S. -C. Wu[6], Y. P. Qi[6], L. X. Yang[2], M. X. Wang[3], Y. Sun[6], N.B.M. Schröter[1], H. F. Yang[1,7], L. M. Schoop[8], Y. Y. Lv[9], J. Zhou[9], Y. B. Chen[9], S. H. Yao[9], M. H. Lu[9], Y. F. Chen[9], C. Felser[6], B. H. Yan[3,6], Z. K. Liu[3]† and Y. L. Chen[1,2,3]†

[1]*Department of Physics, University of Oxford, Oxford, OX1 3PU, UK*
[2]*State Key Laboratory of Low Dimensional Quantum Physics, Department of Physics, Tsinghua University, Beijing 100084, P. R. China*
[3]*School of Physical Science and Technology, ShanghaiTech University and CAS-Shanghai Science Research Center, Shanghai, P. R. China*
[4]*Advanced Light Source, Lawrence Berkeley National Laboratory, Berkeley, CA 94720, USA*
[5]*Accelerator Laboratory, POSTECH, Pohang 790-784, Korea*
[6]*Max Planck Institute for Chemical Physics of Solids, D-01187 Dresden, Germany*
[7]*State Key Laboratory of Functional Materials for Informatics, SIMIT, Chinese Academy of Sciences, Shanghai 200050, P. R. China*
[8]*Max Planck Institute for Solid State Research, 70569 Stuttgart, Germany*
[9]*National Laboratory of Solid State Microstructures, School of Physics and Department of Materials Science and Engineering, Nanjing University, Nanjing, Jiangsu 210093, China*

*These authors contributed equally to this work.

†Correspondence should be addressed to: *liuzhk@shanghaitech.edu.cn*, *Yulin.Chen@physics.ox.ac.uk*

**Topological Dirac semimetals (TDSs) represent a new state of quantum matter recently discovered that offers a platform for realizing many exotic physical phenomena. A TDS is characterized by the linear touching of bulk (conduction and valance) bands at discrete points in the momentum space (i.e. 3D Dirac points), such as in $Na_3Bi$ and $Cd_3As_2$. More recently, new types of Dirac semimetals with robust Dirac line-nodes (with non-trivial topology or near the critical point between topological phase transitions) have been proposed that extends the bulk linear touching from discrete points to 1D lines. In this work, using angle-resolved photoemission spectroscopy (ARPES), we explored the electronic structure of the non-symmorphic crystals MSiS (M=Hf, Zr). Remarkably, by mapping out the band structure in the full 3D Brillouin Zone (BZ), we observed two sets of Dirac line-nodes in parallel with the $k_z$–axis and their dispersions. Interestingly, along directions other than the line-nodes in the 3D BZ, the bulk degeneracy is lifted by spin-orbit coupling (SOC) in both compounds with larger magnitude in HfSiS. Our work not only experimentally confirms a new Dirac line-node semimetal family protected by non-symmorphic symmetry, but also helps understanding and further exploring the exotic properties as well as practical applications of the MSiS family of compounds.**

Topological Dirac semimetals (TDSs) are novel topological quantum materials that were recently discovered and intensively investigated for their rich physics and broad application potential [1-7]. Different from topological insulators (TIs), TDSs are characterized by the linear touching of bulk (instead of surface) bands. In the case of point touching, TDSs can naturally host 3D Dirac fermions, making them the 3D analogue of graphene with many exquisite transport properties, including high mobility charge carriers and large magneto-resistance [3,4]. Typically, in currently known TDSs (e.g. $Na_3Bi$ [1,3] and $Cd_3As_2$ [2,4-7]), the 3D Dirac fermions are protected by time-reversal symmetry and inversion-symmetry. With further symmetry breaking, these TDSs can evolve into more interesting topological states, e.g. to topological Weyl semimetals (TWSs) if time-reversal or inversion symmetry is broken [8-10]. TWSs represent another intensively investigated topological state with many interesting and unusual properties, such as unique surface Fermi arcs and the chiral magnetic effect [11-13].

More recently, a new type of TDS has been predicted in crystals with non-symmorphic symmetry (such as glide planes or screw axis) that induces a band degeneracy at the Brillouin zone boundary, leading to 1D Dirac line-nodes (instead of discrete Dirac points discussed above) [14]. The degeneracy at the line-nodes is protected by non-symmorphic symmetry and robust against perturbations, including spin-orbit coupling (SOC), which breaks the degeneracy along "nodal-rings" in all previously discovered TDSs and TWSs (with the exception of the discrete Dirac or Weyl nodes) [10]. Moreover, comparing to the discrete 3D Dirac points, one-dimensional Dirac line-nodes are a more significant feature in the band structure (Fig. 1(a)) and can have stronger contribution to the physically observable properties of these materials (e.g. in electric transport) [15].

Although topological Dirac line-node semimetals (TDLS) have been predicted in various compounds, only a few of them have been experimentally studied, such as the *M*SiS (*M*=Zr, Hf) family, $PbTaSe_2$ and $PtSn_4$. Among them, $PbTaSe_2$ is inversion asymmetric and hosts "nodal rings" protected by reflection symmetry under SOC. However, the various surface state bands interfering with the bulk features made the identification of nodal rings difficult [16]. $PtSn_4$ has broken nodal-arcs and its topological character is still under investigation [17], hence is also not a simple system to study. Due to its relatively simple electronic band structure, the *M*SiS family naturally becomes a suitable system for the investigation of TDLS. Up to date, although recent ARPES reports on ZrSiS [18-20] and the related compound ZrSnTe [21] have shown that bulk bands with linear dispersion exist (and the Dirac line-nodes was suggested in [18, 19]), there has not yet been a systematic study in the full 3D BZ to demonstrate the Dirac line-nodes and its dispersion in the momentum space. It therefore becomes the goal of this work to clearly illustrate the Dirac line-nodes and their detailed dispersions, discuss the TDLS nature, as well as to study the spin-orbit effect in different compounds of this family of materials.

In this investigation, by making use of high-resolution angle-resolved photoemission spectroscopy (ARPES) with broadly tunable photon energy, we systematically studied the electronic structure of two materials (HfSiS and ZrSiS) of the *M*SiS family compounds. By carrying out broad range momentum space and photon energy dependent measurements, we surveyed multiple BZs along all three momentum directions and were able to identify the electronic bands originating from both the bulk and surface states, which can be well reproduced by our (and previous) *ab-initio* calculations [18-20]. Moreover, by tracking the Dirac band crossing along different high symmetry directions in the BZ, we clearly observed the Dirac line-nodes along the X-R and M-A directions (see Fig. 1(d,e), red lines) and the lifting of the Dirac band degeneracy in other directions due to the SOC (Fig. 1(d,e), green lines). The observation of the Dirac line-nodes in HfSiS and ZrSiS thus proves the Dirac line-nodes semimetal nature of *M*SiS family of compounds, which have demonstrated many intriguing physical properties (e.g. high mobility electrons and "butterfly" shaped titanic angular magnetoresistance ~180000% [15,22-25] ) and hints its criticality to a topological non-trivial phase and thus can be an ideal platform for future investigations and applications.

Due to the larger SOC in HfSiS (compared to ZrSiS), we first focus on this compound. The crystal structure of HfSiS is illustrated in Fig. 1(b), clearly showing the Si-Hf/S-Hf/S-Si quadruple layer structure (PbFCl structure type in the tetragonal P4/nmm space group (No. 129), with lattice constants *a=b=3.52Å, c=8.00Å)*. Square nets of Si form glide planes in the crystal, while the $C_{2x}$ ($C_{2y}$) axis of the lattice are screw axis, thus preserving two kinds of non-symmorphic symmetry. High quality single crystals were synthesized by iodine vapor transport (see Methods for details). After cleaving, large (mm scale) flat and shiny surfaces are exposed (Fig. 1(c)(i)), which is ideal for ARPES measurements. The high quality of the crystal can be verified by the X-ray diffraction (XRD) measurements (Fig. 1(c)(ii-iv)); and the core level photoemission spectrum clearly shows the characteristic $Hf_{4f}$, $S_{2p}$ and $Si_{2p}$ peaks (Fig. 1(c)(v)). The *ab-initio* calculated bulk band structure along high symmetry directions is shown in Fig. 1(e), where all the bands at X and M points are predicted to be at least doubly degenerate as the glide mirror symmetry $\{M_z|\frac{1}{2},\frac{1}{2}\}$ protects the band degeneracy at the X point and the screw axis symmetry $\{C_{2x}|\frac{1}{2},0\}$ protects the band degeneracy at X and M point in the 2D non-symmorphic lattice within P4/nmm group [14]. Extending to 3D, all lines along $k_z$ (R-X, A-M) still preserve the same symmetries with the addition of inversion symmetry, thus forming degenerate lines, including the Dirac line-nodes formed from the crossing of two Dirac-like bands (Fig. 1(d,e)). The broad Fermi surface mapping in Fig. 1(f) covering multiple BZs confirmed the high quality (001) cleavage plane, which allows us to carry out the study below.

We then investigated the detailed electronic structure of HfSiS within a BZ with high-momentum and energy resolution, and an example is shown in Fig. 2. From the 3D plot of the overall band structure (Fig. 2(a)) and the constant energy contours (Fig. 2(b)(i-iv)), one could clearly identify two sets of features. One is the large diamond shaped Fermi surface (FS), forming delicate structures connecting four $\bar{X}$ points; and the other refers to the two concentric ellipses centred at each $\bar{X}$ point. Both features can be well reproduced by our *ab-initio* calculations and previous reports [18-20] (Fig. 2(b)(v-viii)), which also helped us to identify their bulk (the diamond shape FS feature) and surface (the ellipses) origin, respectively (more discussion and the photon energy dependent ARPES measurements can be found below and in the Supplemental Materials).

For the concentric surface state (SS) FSs around the $\bar{X}$ points (see Fig. 2(b, c)), one can see that they are formed by a pair of Rashba-split bands (the splitting is most apparent along $\bar{M} - \bar{X} - \bar{M}$, see Fig. 2d(iii)), indicating the large SOC in HfSiS. The constant energy contours of this set of SS keep shrinking in size and eventually terminate at $E_b \approx 0.4$ eV. Meanwhile, the linear bulk band dispersions (Fig. 2b(iii) and Fig. 2d(i)) form a Dirac point ($E_b \approx 0.6$ eV) at the $\bar{X}$ point. We note that the slight broadening of the bulk band in Fig. 2d(i) is due to the $k_z$ broadening effect [26], as can be clearly seen in the *ab-initio* calculation (Fig. 2d(ii)) where the overlap of dispersions with different $k_z$ momentum (red curves) cause slight broadening. The excellent overall agreement between the experiments (Fig. 2d(i,iii) and e(i)) and the calculation (Fig. 2d(ii,iv) and e(ii)) also allowed us to identify the additional SSs (green curves in Fig. 2d(ii,iv) and e(ii)) from the bulk states (more discussion about BS and SS at $\bar{X}$ can be found in the Supplemental Materials), which simplified our investigation of the bulk states and the identification of the bulk Dirac line-node.

Next we examine the dispersion of the Dirac node along different paths in the BZ. We first study the dispersion along X-R directions (Fig. 3a) by conducting photon energy dependent ARPES (please see supplemental materials for more details on the determination of the $k_z$ momentum). From a series of measurements at different $k_z$ along $(0, 0, k_z) - (\pi/a, 0, k_z)$ direction (see Fig. 3b for a schematic illustration and Fig. 3(c, d)) for sample measurements), we clearly observed that the Dirac node formed by linear band touching disperses with $k_z$ and result in a line-node along X-R direction. With measurements of fine $k_z$ step, the dispersion of the X-R line-node could be reconstructed (by tracking the positions of the DP positions), as illustrated in Fig. 3d(v), which shows good agreement with our calculations (solid red curve).

Similar to the X-R line-node, the predicted bulk band line-node along the M-A direction was also observed (causing by bulk Dirac crossings, see Fig. 1e and Fig. 3e(i-iv)), though with much less dispersion (Fig. 3e(v)). The observation of the Dirac line node-along X-R and M-A unambiguously confirms that HfSiS is a Dirac line-nodes semimetal.

Similarly, our measurement on ZrSiS have also obtained similar results on the Dirac line-nodes (see Fig. 4), with the Dirac line node along X-R and M-A directions, but with slightly weaker dispersion along the R-X-R direction (Fig. 4(c)).

Due to their similarity in their crystal structure, the comparison between the band structure of HfSiS and ZrSiS can provide important information about the effect of SOC in this family of Dirac line-nodes semimetals; one clear example is the effect on the surface state around the $\bar{X}$ point, where the two concentric pockets with a distinct separation can be seen in HfSiS (Fig. 5b(i-ii), the Rashba parameter estimated is 3.1eVÅ); while the separation, if any, is smaller than our experimental resolution in ZrSiS (Fig. 5b(iii-iv)).

In addition to the surface states, the SOC also plays an important role in causing large bulk band splitting. As an example, along the Γ-X direction (Fig. 5c), two bulk bands at high binding energy showing clearly larger splitting in HfSiS (Fig. 5c(i)) than that in ZrSiS (Fig. 5c(iii)), agreeing with our calculations (Fig. 5c(ii, iv)) that also predict larger splitting in HfSiS. The effect of the SOC can also be seen at the crossing between the two Dirac-like bands halfway between $\bar{\Gamma} - \bar{M}$, where the opening of a small bandgap is also larger in HfSiS (~80 meV, see Fig. 4d(i)) than that in ZrSiS, (~15meV, see Fig. 5d(iii)). This is in contrast to the degenerate Dirac line nodes along X-R and M-A in the BZ, which are protected by the non-symmorphic symmetry in both compounds.

Our systematic study on the electronic structures and the observation of their Dirac line-nodes in $M$SiS ($M$=Zr, Hf) family of compounds along the X-R and M-A directions (while the degeneracy is lifted in other directions) in the BZ clearly establish that these compounds are Dirac line-nodes semimetals protected by non-symmorphic symmetry. The detailed characterization of the electronic structures of this family of compounds and the effect of the SOC will not only help us understand their intriguing transport properties even if the Dirac point observed is ~0.6-0.8eV below $E_F$ (e.g. the Dirac-like fermion contribution seen in the de Haas–van Alphen quantum Oscillations [23], the non-trivial Berry phase and magnetoresistance[20, 22, 27], as well as the distinct electron correlation effects due to the different energy scaling of the density of electronic states [28]), but also provide a solid base for designing and realizing exotic topological quantum phases and possible future applications with these materials. Especially, by tuning the Dirac point of the MSiS materials to the $E_F$ vicinity, these materials may lead to more interesting discovery in the transport properties from the Dirac line-nodes semimetals.

Finally, although the theoretical calculations so far suggest that these two compounds do not have non-trivial topology, there are gathering transport and photoemission evidences which strongly suggest a topologically non-trivial phase in the MSiS family [19-25, 27, 29]. In addition, the family compounds are close to the border of the topological phase transition; and many materials with the same crystal structures have rich topologically non-trivial phases, such as (non-symmorphic symmetry protected) topological insulators and topological semimetals [30-31]. The application of external influence such as mechanical strain or chemical substitution further opens up the opportunity to drive the topological phase transition of MSiS materials. Thus we conclude the MSiS material is at least a topologically critical line-nodes semimetal and believe the study of these materials will be important for the systematic study of the topological phase diagram and the associated topological phase transitions of the MSiS and relevant compounds.

As we are preparing the manuscript, we noticed an ARPES report has also addressed the surface band splitting due to the stronger SOC strength in HfSiS [29].

## APPENDIX: MATERIALS AND METHODS

**1. Sample synthesis:**

Single crystals of HfSiS were grown via iodine ($I_2$) vapor transport in a two steps process. At first, the polycrystalline HfSiS powder was synthesized using high-purity elemental Hf piece, Si piece, S powder in a process described elsewhere [32]. Then the polycrystalline powder together with iodine were sealed in a quartz tube under vacuum. The quartz tube was then kept in a gradient furnace for 10 days with the powder at 1100°C, while the cooler end at 1000°C. Shiny rectangular plate like crystals were obtained at the cooler end for ARPES measurements.

**2. Angle resolved photoemission spectroscopy:**

ARPES measurements were performed at the beamline I05 of the Diamond Light Source (DLS) and beamline 13U of the National Synchrotron Radiation Laboratory (NSRL), Hefei, both equipped with Scienta R4000 analyzers. The measurement sample temperature and pressure were 10 K and lower than $1.5\times10^{-10}$ Torr, respectively. The angle resolution was $0.2°$ and the overall energy resolutions were better than 15meV. The samples were cleaved in situ along the (001) plane.

**3. *Ab-initio* calculations:**

The density functional theory (DFT) calculations have been performed by using the Vienna ab initio simulation package (VASP) with the projected augmented wave method (PAW) [33]. The exchange and correlation energy was considered in the generalized gradient approximation (GGA) with Perdew-Burke-Ernzerhof (PBE) based density functional [34]. The energy cutoff was set to 259 eV. Experimental lattice constants were used for two compounds in all the calculations. A (001) slab model of five bulk units thick (40 Å) was constructed with the periodicity in the xy plane and a vacuum layer of 20 Å in the z direction to simulate the (001) surface. The slab band structure was projected to the atomic orbitals of top one unit cell, to identify the surface states.

**Acknowledgements:**

*Y.L.C acknowledges the support from the EPSRC (UK) grant EP/K04074X/1, a DARPA (US) MESO project (no. N66001-11-1-4105) and Hefei Science Center CAS (2015HSC-UE013). C.F. acknowledges the financial support by the ERC Advanced Grant (No. 291472 "Idea Heusler"). C.C acknowledges the support of the China Scholarship Council-University of Oxford Scholarship. J. J. acknowledges the support of the NRF, Korea through the SRC center for Topological Matter (No. 2011-0030787). H.F.Y acknowledges the finacial support from the Bureau of Frontier Sciences and Education, Chinese Academy of Sciences. N.B.M.S acknowledges the support from Studienstiftung des deutschen Volkes.*

**Author Contributions:**

*Y.L.C conceived the experiments. C.C and Z.K.L carried out ARPES measurements with the assistance of J.J, L.X.Y and M.X.W. Y.P.Q and S.H.Y synthesized and characterized the bulk single crystals. S.C.W, Y.S, J.Z. and B.H.Y performed ab-initio calculations. All authors contributed to the scientific planning and discussions.*

**Competing financial interests:**

*The authors declare no competing financial interests.*

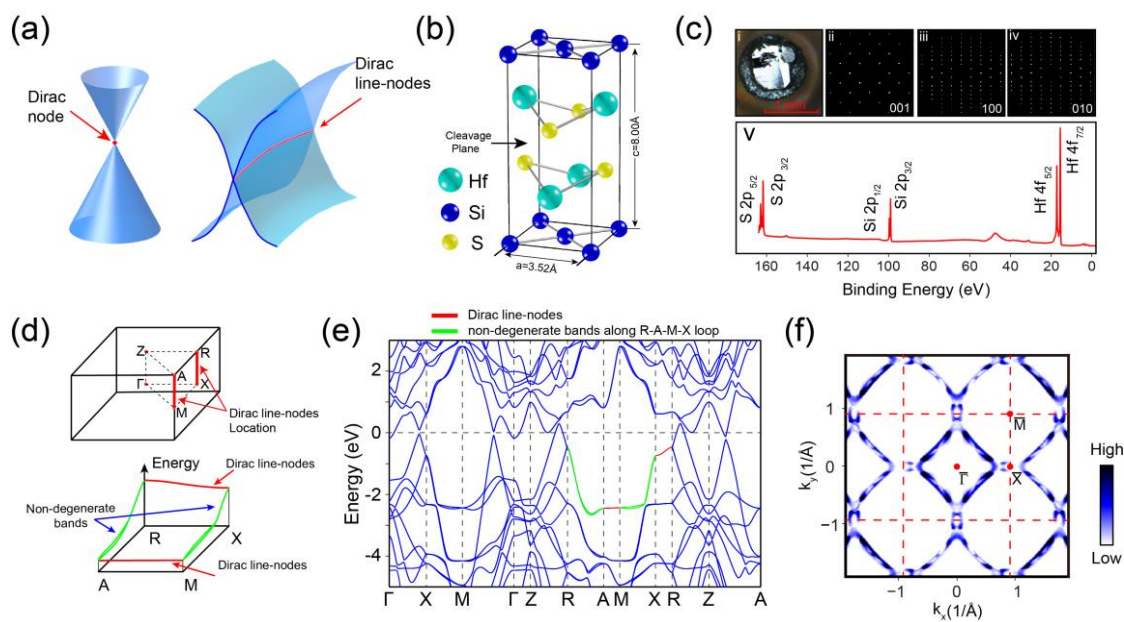

**FIG. 1 (color online). Dirac line-nodes and characterization of HfSiS single crystals. (a),** Schematic illustration of Dirac Node and Dirac line-nodes in the momentum space. **(b),** Crystal structure of the HfSiS, showing the stacking of Si and Hf-S planes. Lattice constant and most probable cleavage plane is indicated. **(c),** (i) Photograph of the HfSiS crystal cleaved along <001> direction, (ii-iv) XRD pattern of the (001), (100) and (010) surface of HfSiS and (v) Core level spectrum of HfSiS showing characteristic $S_{2s}$, $Si_{2p}$ and $Hf_{4f}$ core level peaks. **(d),** Illustration of the Dirac line-nodes location (upper panel) and the dispersion of the line-nodes (lower panel). **(e),** Calculation results of the bulk band structure of HfSiS along high-symmetry directions. Green lines indicate the positions where the bands are non-degenerate while the red lines indicate the Dirac line-nodes. **(f),** Broad range photoemission spectral intensity map of the Fermi-surface covers 9 BZ's, showing the correct symmetry and the characteristic FS.

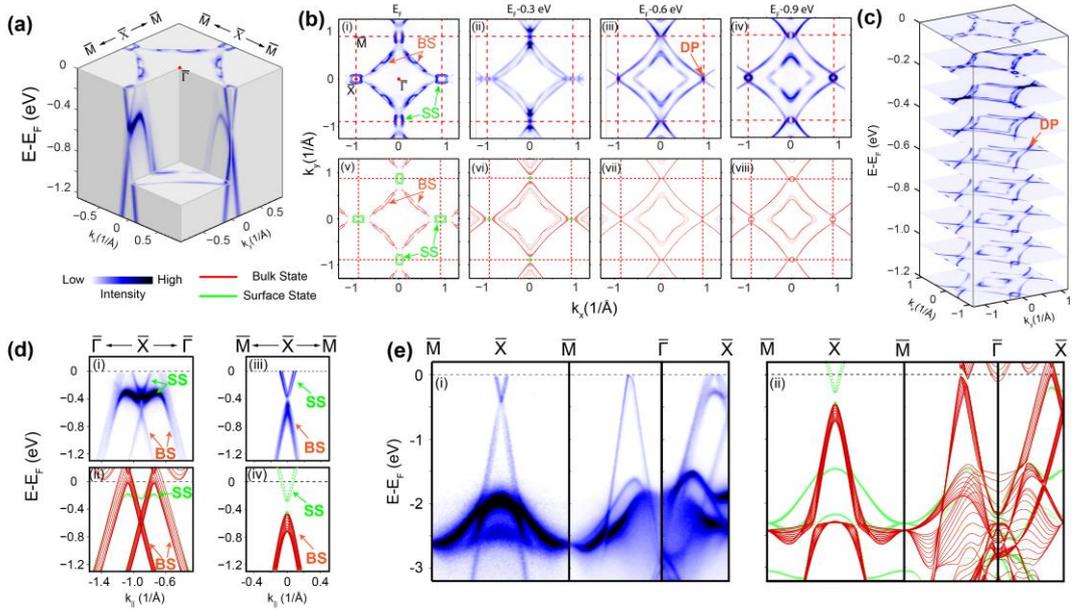

**FIG. 2 (color online). General electronic structure of HfSiS. (a),** 3D intensity plot of the photoemission spectra centered around $\bar{\Gamma}$. **(b),** Top row: Photoemission spectral intensity map showing the constant energy contours of bands at $E_B$=0 (i), 0.3 (ii), 0.6 (iii) and 0.9 eV (iv), respectively. Bulk state (BS), surface state (SS) and Dirac point (DP) are labelled. Bottom row (v)-(viii): Corresponding calculated constant energy contours at the same binding energies as in experiments above. **(c),** Stacking plots of constant-energy contours in broader binding energy range show the band structure evolution. The position of DP is labelled. **(d),** High symmetry cut along the $\bar{\Gamma} - \bar{X} - \bar{\Gamma}$ (i) and $\bar{M} - \bar{X} - \bar{M}$ (iii) directions near $E_F$, together with their corresponding calculated bandstructure (ii,iv). BS and SS are labelled on the data and represented by red and green color in the calculation. **(e),** Broad range high symmetry cut along the $\bar{M} - \bar{X} - \bar{M} - \bar{\Gamma} - \bar{X}$ directions (i) and corresponding calculated bandstructure (ii). BS and SS are represented by red and green color in the calculation.

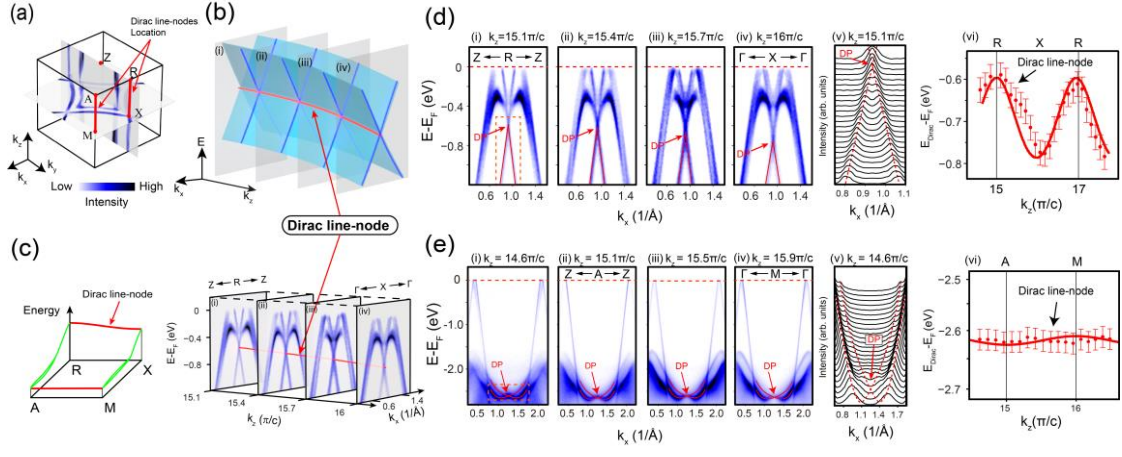

**FIG. 3 (color online). Dirac line-nodes in HfSiS. (a),** Illustration of the location of Dirac line-nodes in HfSiS. The 3D Brillouin Zone also shows the constant energy contours in the $k_y$-$k_z$ plane (with $k_x=0$) and $k_x$-$k_y$ plane (with $k_z=0$) with energy around the DP. **(b),** Schematic of the bandstructure hosting a Dirac line-node along the $k_z$ direction. Grey planes represent locations where each cut in **(c)** is measured and blue lines represent the Dirac dispersion we would observe in each cut. The Dirac line-node is labelled by the red curve. **(c),** Left panel, Schematic of the dispersion of the line-node. Right panel, stack plot of the four cuts along $(k_x,k_y)=(0,0)$ to $(k_x,k_y)=(0,\pi/a)$ with $k_z$ values ranging from 15 to 16 $\pi/c$. Red curve indicate the Dirac line-node. **(d),** (i)-(iv) Detailed analysis of the four cuts shown in **(c)** with the linear fitting results of the Dirac bands marked by red lines and DP labelled. (v) MDC plot for the region around the Dirac point (indicated by the dashed box in (i)). (vi) The fitting DP positions at various $k_z$ values, the red curve indicates the calculated dispersion along the R-X-R direction. **(e),** (i)-(iv) Plot and band fitting result of four representing cuts along $(k_x,k_y)=(0,0)$ to $(k_x,k_y)=(\pi/a,\pi/a)$ with $k_z$ values ranging from 14.6 to 16 $\pi/c$. Fitted parabolic dispersions and DP positions are labelled. (v) MDC plot for the region around the Dirac point (indicated by the dashed box in (i)). (vi) The fitting DP positions at various $k_z$ values, the red curve indicates the calculated dispersion along the A-M-A direction.

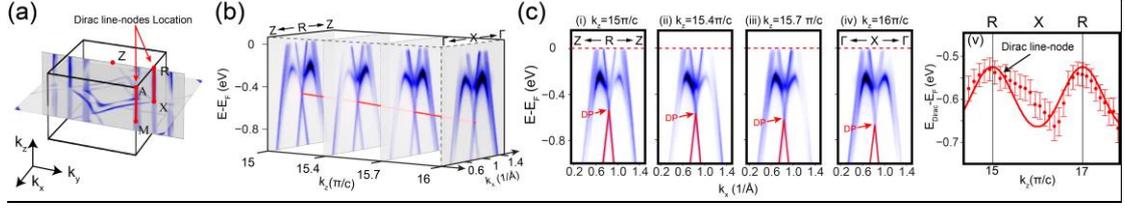

**FIG. 4 (color online). Dirac line-nodes in ZrSiS. (a),** Illustration of the location of Dirac line-nodes in ZrSiS. The 3D Brillouin Zone also shows the constant energy contours in the $k_y$-$k_z$ plane (with $k_x=0$) and $k_x$-$k_y$ plane (with $k_z=0$) with energy around the DP. **(b),** Stack plot of the four cuts along $(k_x,k_y)=(0,0)$ to $(k_x,k_y)=(0,\pi/a)$ with $k_z$ values ranging from 15 to 16 $\pi/c$. Red curve indicates the Dirac line-node. **(c),** (i)-(iv) Detailed analysis of the four cuts shown in **(b)** with the linear fitting results of the Dirac bands marked by red lines and DP labelled. (v) The fitting DP positions at various $k_z$ values, the red curve indicates the calculated dispersion along the R-X-R direction.

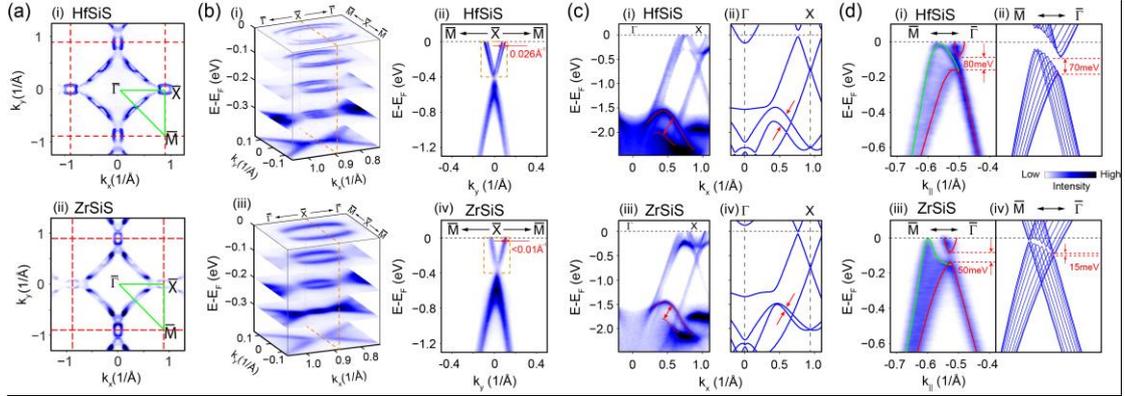

**FIG. 5 (color online). Effect of Spin-Orbit Coupling (SOC) on the electronic structure of HfSiS and ZrSiS. (a),** Constant energy contour on $E_F$ of HfSiS (i) and ZrSiS (ii). Green lines indicate the cut directions in (b)-(d). **(b),** (i-ii) 3D intensity plot of the photoemission spectra centered around $\bar{X}$ and the high symmetry $\bar{\Gamma}-\bar{X}$ cut of HfSiS. (iii-iv) Same plot for ZrSiS. The SS splitting in both materials are estimated and labelled. **(c),** Comparison of the high symmetry Γ-X cut together with their bandstructure calculation of HfSiS (i-ii) and ZrSiS (iii-iv). The band splitting in both materials are labelled. **(d),** Comparison of the high symmetry $\bar{M}-\bar{\Gamma}$ cut together with their bandstructure calculation of HfSiS (i-ii) and ZrSiS (iii-iv). In the spectrum, red curve labels the bandstructure at $k_z=\pi$ while the green curve labels the

bandstructure from other $k_z$ values. The band gap between the electron and hole pockets are estimated and labelled. In the calculation results, the band structure of all $k_z$ values are plot together to reflect the uncertainty of $k_z$ values each cut covers. The band splitting between the electron and hole pockets at $k_z = \pi$ are labelled.